\documentclass[aps,pra,twocolumn,groupedaddress]{revtex4}

\usepackage{amsmath}
\usepackage{amssymb}
\usepackage{mathtools}
\usepackage{gensymb}
\usepackage{bbm, dsfont}
\usepackage[utf8]{inputenc}
\usepackage{bbold}
\usepackage{color}

\usepackage{float}
\usepackage[capitalize]{cleveref}

\begin{document}

\title{Blind steering in no-signalling theories}

\author{E. Zambrini Cruzeiro$^{1}$, N. Gisin$^{2,3}$, S. Popescu$^{4,5}$}

\affiliation{$^1$\textit{Laboratoire d'Information Quantique, CP 225, Université Libre de Bruxelles, 1050 Bruxelles, Belgium}\\
$^2$\textit{Group of Applied Physics (GAP), University of Geneva, 1211 Geneva, Switzerland}\\
$^3$\textit{Schaffhausen Institute of Technology - SIT, Geneva, Switzerland}\\
$^4$\textit{School of Physics, University of Bristol, Tyndall Avenue, Bristol BS8 1TL, U.K.}
\\
$^5$\textit{Institute for Theoretical studies, ETH Zürich, Switzerland}}

\begin{abstract}
Steering is a physical phenomenon which is not restricted to quantum theory, it is also present in more general, no-signalling theories. Here, we study steering from the point of view of no-signalling theories. First, we show that quantum steering involves a collection of different aspects, which need to be separated when considering steering in no-signalling theories. By deconstructing quantum steering, we learn more about the nature of the steering phenomenon itself. Second, we introduce a new concept, that we call "blind steering", which can be seen as the most basic form of steering, present both in quantum mechanics and no-signalling theories.
\end{abstract}


\maketitle

\section{Introduction}

Entanglement is one of the key ingredients of quantum mechanics. In 1935, A. Einstein, B. Podolsky and N. Rosen (EPR) gave an argument for the incompleteness of quantum mechanics \cite{Einstein1935}, based on measurements of position or momentum on a two-party entangled pure state. In such a scenario, the correlations of the entangled state allow one party to guess the measurement outcomes of the other party, if both perform the same measurement. This seemingly absurd idea shook the most prominent physicists’ minds at the time, including E. Schrödinger, who soon after described formally the ``spooky action at a distance'' introduced by EPR: one party can steer the state of the other party into an eigenstate of position or momentum \cite{Schrodinger1936}. Much more recently \cite{Wiseman2007}, steering was shown to represent a notion in between entanglement and nonlocality. Several forms of steering have been presented over the years \cite{Schrodinger1936,Hadjisavvas1981,Gisin1989,Hughston1993,Wiseman2007,Sainz2015,Sainz2020}. In each of them, a different question is asked. It is also a natural scenario to discuss quantum information protocols where trust in the measurement device of one of the parties is not required; these are called one-side device-independent protocols \cite{Cavalcanti2016}. Steering has helped solve open problems in quantum information theory, notably finding counterexamples to the Peres conjecture \cite{Pusey2013,Skrzypczyk2014,Moroder2014,Vertesi2014}. Steering was also shown to be tightly related to the concept of joint measurability of generalized measurements \cite{Quintino2014,Uola2014,Uola2015,Kiukas2017}. Even more recently, a strong link between contextuality and steering has been demonstrated \cite{Tavakoli2020}. For a comprehensive review of steering, see \cite{Uola2020}.

Similar to the case of non-locality, which was first discovered in the context of quantum mechanics, and then shown to be a phenomenon in itself, which could be present in various other theories, steering could also be viewed in and by itself, independent of quantum mechanics as shown in \cite{Chiribella2010,Barnum2013,Bae2012,Stevens2014,Banik2015,Sainz2015,Plavala2017,Jenvcova2018,Hoban2018,Sainz2020}. This is the point of view we take in our present paper.  

We present two main results. First, we show that in quantum mechanics "steering" involves a collection of different aspects; going to more general theories we learn that these aspects that were collated together in quantum steering, need to be separated. This way, by de-constructing quantum steering, we learn more about the nature of the steering phenomenon itself, as well as about the particular properties of quantum mechanics. Second, we propose a new concept of steering, that we call "blind steering". Although it is motivated by the study of steering for nonlocal no-signalling boxes with correlations stronger than those possible in quantum mechanics, it can also be considered in a quantum mechanical context. 

\section{Steering with boxes}

\subsection{Deconstructing the GHJW theorem}

An early question, going to the very basis of steering was raised in \cite{Gisin1989}. Consider any two ensembles $\mathcal{E}_0$ and $\mathcal{E}_1$ corresponding to the same density matrix $\rho$. Is it possible to construct a particular joint state $\Phi_{AB}$, distributed between two observers, Alice and Bob, such that, by appropriate measurements Bob can steer Alice's state into either the first or second ensemble? The question was then answered positively: All that is needed is to start with a pure state $\Psi_{AB}$ which is such that (i)  Alice's reduced density matrix equals $\rho_A=\rho$, the density matrix of interest, and (ii) Bob's subsystem has Hilbert space dimension as large as the number of elements in the largest ensemble. 

A subsequent result, \cite{Hughston1993}, significantly extended the above: It was shown that {\it any} given pure state $\Psi_{AB}$ that has reduced density matrix $\rho_A=\rho$ can be used for steering regardless 
of the dimension of the Hilbert space of Bob's subsystem. Moreover, Bob could not only steer between two ensembles $\mathcal{E}_1$ and $\mathcal{E}_2$ consistent with $\rho_A$ but to {\it any} of the infinite number of such ensembles.  The results in \cite{Gisin1989} and \cite{Hughston1993} are known as the GHJW Theorem \cite{Sainz2015,Mermin1999}.

The crucial element that enabled the stronger result \cite{Hughston1993} is the observation that Bob's subsystem doesn't need to have Hilbert space dimension as large as the number of elements in the largest ensemble because Bob could simply use a separate ancilla, perform a joint measurement on his subsystem and the ancilla, and achieve the same effect as when Bob's subsystem would have had larger dimension. (Technically, Bob can perform on his subsystem any POVM). 

Quantum mechanically the result of \cite{Hughston1993} supersedes the result on \cite{Gisin1989}. Coming to no-signalling boxes, the situation turns out to be different: Indeed, while it is possible to have non-local correlations stronger than those possible in quantum mechanics, the {\it dynamics} turns out to be far more restricted than in quantum mechanics. In particular, it is in general not possible for Bob to perform such a joint measurement between his box and an ancillary one \cite{Short2006,Barrett2007,Short2010}. In other words, with no-signalling boxes, general POVMs cannot be reproduced. The generalisation of \cite{Hughston1993} to NS boxes is thus not possible. However, as we show below, the result of \cite{Gisin1989} can be generalised.

The significance of this result, as far as we see it, is the following: The fundamental meaning of steering is the possibility to remotely prepare one out of two arbitrary ensembles, constrained only by the requirement of no signalling. This may require special preparations. That in some theories, such as quantum mechanics, the special preparations are less stringent, and more powerful, is an additional property of those theories which has nothing to do with steering itself. Indeed, the way of constructing POVM's by von Neumann measurements on a system plus ancilla is {\it not} a property of steering. It impacts on steering, but it is not a steering property. Hence, this result exposes the essence of steering, separating it from additional, unrelated, phenomena. 

\subsection{Box ensembles}
\label{subsec:bwensembles}

In no-signalling (NS) theories \cite{Popescu1994,Barrett2007}, the basic objects are ``boxes'' which could be shared by many parties. A box is an input-output device. The probability distribution of the outcomes given the inputs can be thought of the ``state'' of the box.

In the case where only one party is using the box, we call it a ``local box''. Such a local box can be defined for any finite number of inputs $x=0,\dots,X-1$, and outputs $a=0,\dots, A-1$. Given a local probability distribution $p(a|x)$ for Alice, an ensemble $\mathcal{E}$ realizing $p(a|x)$ is a set of $n$ boxes, along with probabilities $w_j$ associated to each box, where $j=0,\dots ,n-1$. We restrict to ensembles of deterministic local boxes, the analogue of ensembles of pure states in quantum mechanics, without loss of generality. We associate to each deterministic local boxes a probability distribution $p_j(a|x)$ which can be written
\begin{equation}
p_j(a|x)=\delta_{a,f_j(x)}
\end{equation} 
where $f_j$ is the function describing the jth deterministic strategy, $j=0,\dots ,n-1$.

An ensemble realizing $p(a|x)$ is thus defined in the following way
\begin{equation}
\mathcal{E} := \left\{w_j,p_j(a|x)\right\}_{j=0}^{n-1}
\end{equation}
such that
\begin{equation}\label{eq:3}
p(a|x) = \sum_{j=0}^{n-1} w_jp_j(a|x)
\end{equation}

\subsection{Box steering}

In the theorem that follows, we introduce box steering. By this, we mean that Bob can prepare remotely any of the local boxes contained in two distinct ensembles realizing the same local state $p(a|x)$. These ensembles contain the total number of extremal boxes which can be prepared on Alice's side, i.e. $n=A^X$.

\vspace{0.2cm}

\textit{Theorem 1}: Consider an arbitrary state $p(a|x)$, and two ensembles $\mathcal{E}_0$ and $\mathcal{E}_1$ corresponding to it. Then, there exists a no-signalling state $p(ab|xy)$ such that when Bob performs measurement $y$, Alice obtains ensemble $\mathcal{E}_y$. Crucially, the number of outcomes must be equal to the number of states in $\mathcal{E}_y$. Bob, knowing $y$ and $b$, knows in each individual round which state from the ensemble $\mathcal{E}_y$ Alice has.  

\vspace{2mm}

\textit{proof}: We will show that the no-signalling state
\begin{equation}
p(ab|xy)=w_b^yp_b(a|x)
\end{equation}
allows Bob to prepare the following ensembles through his measurement choice $y$,
\begin{equation}
\mathcal{E}_y=\{w_b^y,p_b(a|x)\}
\end{equation} 

We proceed by showing the steps of the proof.

\begin{itemize}
 \item[i)] The weights $w_b^y$ obey
\begin{equation}
\sum_bw_b^0p_b(a|x)=\sum_bw_b^1p_b(a|x)=p(a|x)
\end{equation}
such that both $\mathcal{E}_0$ and $\mathcal{E}_1$ both correspond to $p(a|x)$. The weights are normalized, $\sum_bw_b^y=1$.
 \item[ii)] We show $p(ab|xy)$ is indeed a probability distribution, because it is positive and normalized,
\begin{equation}
p(ab|xy)\geq 0
\end{equation}
\begin{equation}
\sum_{ab}p(ab|xy)= \sum_bw_b^y\sum_ap_b(a|x)=\sum_bw_b^y=1
\end{equation}
since $\sum_bw_b^y=1$ and by definition of the ensembles.
 \item[iii)] We show $p(ab|xy)$ is no-signalling,
\begin{align}
\begin{split}
 \sum_bp(ab|xy) & =\sum_bw_b^yp_b(a|x)=p(a|x)\\
 \sum_ap(ab|xy) & =p(b|y)
\end{split}
\end{align}
 \item[iv)] When Bob measures, Alice's state becomes $p(a|x;yb)=p_b(a|x)$, where we have separated Alice and Bob's variables using a semicolon. Hence, if we know $y$ and $b$, we know that Alice has the state $p_b(a|x)$.

The probability to get $b$ when Bob measures $y$ is
\begin{equation}
p(b|y)=\sum_ap(ab|xy)=\sum_aw_b^yp(a|x)=w_b^y
\end{equation}
\end{itemize}

\hspace*{\fill} \textsc{q.e.d.}

Note that this proof can be trivially extended to any number of distinct ensembles one wishes to prepare.

We conclude that Bob can steer Alice to any deterministic box of any ensemble, provided that Bob has enough outputs, i.e. $B \geq A^X$. What happens if this is not the case? Can one still steer? Surprisingly in the next section, we answer this question in the affirmative.

\section{Blind steering}

In this section we define a new concept of steering, that we call ``blind steering''. We are motivated to do this due to our analysis of steering for arbitrary non-local boxes, but the concept is general and it applies to any no-signalling theories, in particular to quantum mechanics. 

Consider first the standard steering problem. Let $p(ab|xy)$ be a joint state shared by Alice and Bob. Corresponding to this Alice has a reduced state $p(a|x)$. If this reduced state is not a vertex, it can be represented by many different mixtures of some other of Alice's states, say, of Alice's vertices. Each mixture is a different ``ensemble'', and Bob's task is to steer in between these various ensembles by choosing his measurement appropriately. Furthermore, the result he obtains should tell him which particular vertex in that ensemble has been obtained in each round of the experiment. As we discussed in the previous sections, this task is not possible in general.

Let us change now the problem. Suppose that the joint state $p(ab|xy)$ shared by Alice and Bob is not a vertex of the no-signalling polytope. Then $p(ab|xy)$ itself can be obtained by various mixtures (ensembles) of joint Alice-Bob states, say, of the vertices of the Alice-Bob no-signalling polytope. So let's consider now a three party problem, involving Alice, Bob and a Referee.  

Suppose that the Referee prepares a particular ensemble that corresponds to $p(ab|xy)$. In each round of the experiment that he gives to Alice and Bob one of these states, with the appropriate probability. The referee doesn't inform Alice and Bob what the ensemble is, only to what state $p(ab|xy)$ it corresponds (actually, the Referee doesn't even need to inform them what $p(ab|xy)$ is - given enough rounds of the experiment Alice and Bob can  find it by themselves). 

Bob can then perform, say, one out of two measurements. Depending on which measurement he performs Alice is left in one out of two ensembles. However, suppose that Bob's measurements have a small number of possible outcomes, smaller than the number of constituent states in Alice's local ensemble. Then, Bob, using his knowledge of what measurement he performed, what result he obtained, and what $p(ab|xy)$ is, cannot, in general, infer what state Alice ends up with in each particular round. However, if Bob informs the Referee what measurement he did and what result he obtained, the Referee is able to infer what constituent state Alice holds, thanks to his additional knowledge of the initial two partite ensemble, and the specific constituent two partite state in each round. Hence the Referee could check that steering succeeded.

We call this protocol ``blind steering'' since Bob doesn't know what constituent state he prepared on Alice's site. In some sense, this protocol exhibits the core of what steering is. Bob can steer, even though he himself doesn't know what he steers into. In some sense this is reminiscent of teleportation, where Alice and Bob can teleport an unknown state. 

In the remaining of this section we present an example of blind steering, in the case where Alice and Bob each perform two binary-outcome measurements. 

\begin{figure}[H]
\includegraphics[width=80mm,scale=0.3]{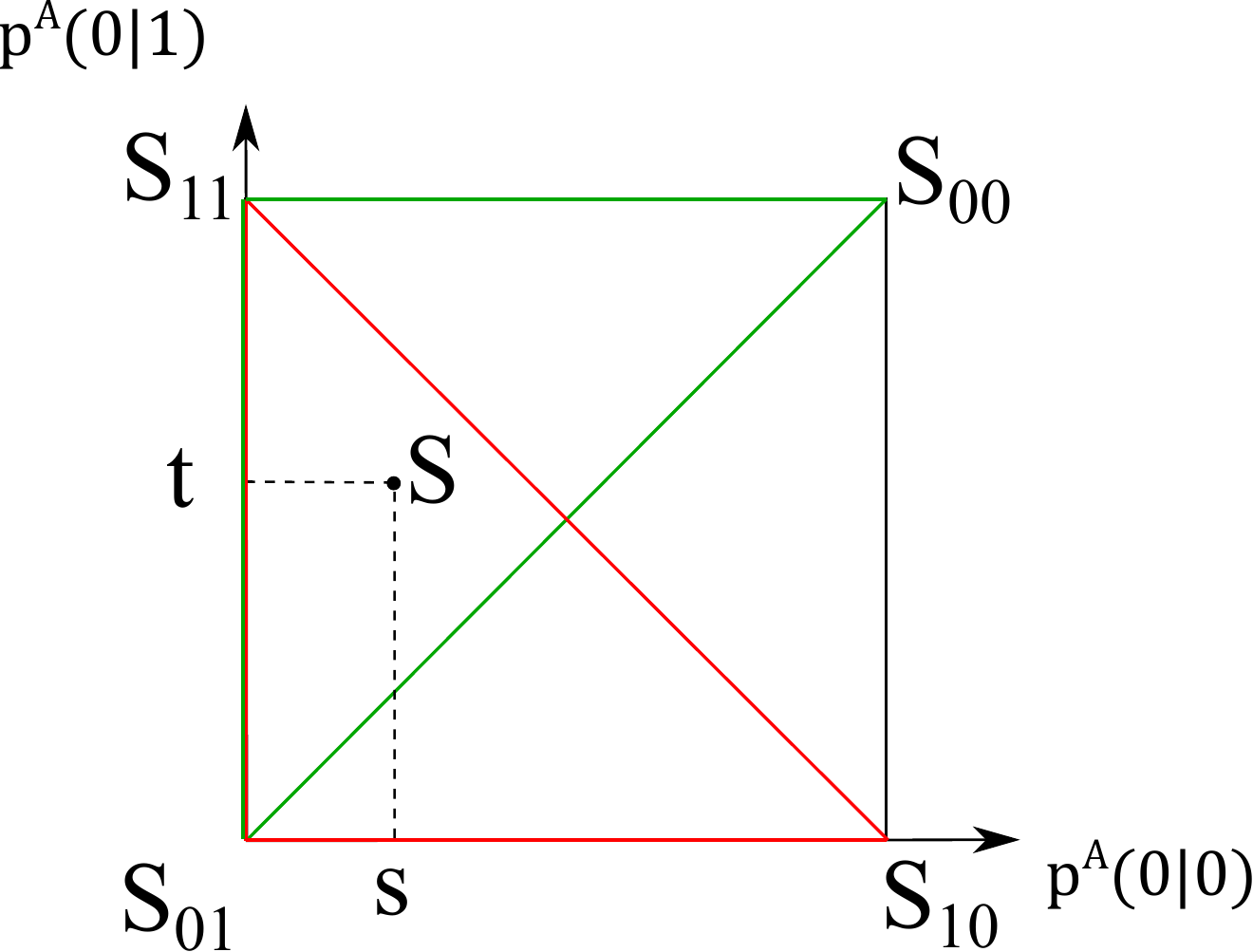}
\caption{Local polytope of Alice, where a particular mixed state $S$ is given coordinates $(s,t)$. The vertices of the local polytope are the local boxes $S_{ij}$. The figure shows that to realize any mixed state $S$ which is not on a diagonal, one can use two sets of three vertices, making two triangles in the local polytope, shown in red and green colors.\label{fig1}}
\end{figure}

When a binary choice of measurements, each with binary outcomes, is considered, each party has four local boxes. In Fig. \ref{fig1}, we depict Alice's local boxes as vertices of the set of states. Alice has a state $S$. It is a mixed state, and can be decomposed in particular into two ensembles composed of three constituent states each. Bob would like to be able to steer between those two ensembles. In particular, for the $S$ state shown in Fig. \ref{fig1}, this means that if $y=0$, an ensemble composed of the boxes $S_{01}, S_{11}, S_{00}$ is prepared. If, on the other hand, Bob selects $y=1$, then he prepares an ensemble composed of boxes $S_{01}, S_{11}, S_{10}$. We will refer to these ensembles as the ``triangle decompositions''. The $S_{\alpha\beta}$ are local boxes, defined as
\begin{equation}
a=\alpha x\oplus\beta
\end{equation}
for Alice, where $\oplus$ denotes addition modulo two. The local boxes of Bob are obtained by simply replacing $x$ with $y$ and $a$ with $b$.

Bob can blind steer with the following protocol. The Referee provides Alice and Bob with a nonlocal ensemble,
\begin{equation}\label{eq:NLbox}
\left\{p_{ijkl},S_{ij}\times S_{kl};q_{\alpha\beta\delta},\mathrm{PR}_{\alpha\beta\delta}\right\}
\end{equation} 
where $p_{ijkl}\geq 0$ $\forall i,j,k,l$, $q_{\alpha\beta\delta}\geq 0$ $\forall \alpha,\beta,\delta =0,1$, and $\sum_{ijkl}p_{ijkl}+\sum_{\alpha\beta\delta}q_{\alpha\beta\delta}=1$. The $\mathrm{PR}_{\alpha\beta\delta}$ are the nonlocal no-signalling extremal strategies, also called PR boxes, defined by the algebraic relation
\begin{equation}
 a\oplus b=(x\oplus\alpha)(y\oplus\beta)\oplus\delta ,
\end{equation}

From the point of view of Alice and Bob, they hold a nonlocal state which is a mixture of the boxes of the nonlocal ensemble: they do not know which decomposition they have. We want to show that the Referee can pick some particular $p_{ijkl}, q_{\alpha\beta\delta}$ to allow the steering into the two triangle decompositions realizing $S$. We will now show what those conditions are. 

Two distinct general local decompositions giving the same state $S$ can be written
\begin{equation}
 S=\sum_{ij}\epsilon_{ij}S_{ij}=\sum_{ij}\eta_{ij}S_{ij}
\end{equation}
where $\epsilon_{ij},\eta_{ij}\geq 0, \forall ij, \sum_{ij}\epsilon_{ij}=\sum_{ij}\eta_{ij}=1$, and the equality holds because the two quantities represent the same state. 

This implies, at the level of the probabilities,
\begin{align}
p(a=0|x=0) &=\epsilon_{00}+\epsilon_{10}=\eta_{00}+\eta_{10}\equiv s\\
p(a=0|x=1) &=\epsilon_{00}+\epsilon_{11}=\eta_{00}+\eta_{11}\equiv t
\end{align}

Note that it is sufficient to specify $p(a|x)$ for one of the outputs because we are dealing with binary outcomes.

We assume that the local state, identified by the coordinates $(s,t)$, is strictly inside one of the triangles delimited in Fig.~\ref{fig1} by colored (green or red) lines. Without loss of generality, we consider the left one, as in the figure. Points within this triangle satisfy both $t\geq s$ and $s+t<1$. We are interested in local ensembles of three constituent states each, therefore we look for the two ensembles $\epsilon_{ij}$ and $\eta_{ij}$ with the largest and smallest possible weights on $S_{00}$, respectively. 

$t\geq s$ implies
\begin{equation}\label{eq:eps}
\epsilon_{00}=s,\epsilon_{01}=1-t,\epsilon_{10}=0,\epsilon_{11}=t-s
\end{equation}

For $\eta$, assuming $s+t<1$,
\begin{equation}\label{eq:eta}
\eta_{00}=0,\eta_{01}=1-s-t,\eta_{10}=s,\eta_{11}=t
\end{equation}

Following a measurement choice $y$ by Bob, we derive the local ensemble on Alice's side.

When $y=0$,
\begin{align}\label{eq:reduced1}
\begin{split}
& \left\{\sum_{b}\left(p_{ij0b}+p_{ij1b}+\frac{1}{2}(q_{0,i,j\oplus b}+q_{1,i,i\oplus j\oplus b})\right),S_{ij}\right\}\\
& = \{P_{ij}+\frac{1}{2}Q_i,S_{ij}\}\\
& = \{\epsilon_{ij},S_{ij}\}
\end{split}
\end{align}
where $P_{ij}=\sum_{kl}p_{ijkl}$ and $Q_{\beta}=\sum_{\alpha\gamma}q_{\alpha\beta\gamma}$.

When $y=1$,
\begin{align}\label{eq:reduced2}
\begin{split}
& \left\{\sum_{b}\left(p_{ij0b}+p_{i,j,1,b\oplus 1}+\frac{1}{2}(q_{0,i\oplus 1,j\oplus b}+q_{1,1\oplus i,j\oplus i\oplus b})\right),S_{ij}\right\}\\
& = \{P_{ij}+\frac{1}{2}Q_{i\oplus 1},S_{ij}\}\\
& = \{\eta_{ij},S_{ij}\}
\end{split}
\end{align}

Combining Eqs.~(\ref{eq:eps}), (\ref{eq:eta}) together with \cref{eq:reduced1,eq:reduced2}, we get the following system of equations

\begin{align}\label{eq:sysfinal}
\begin{split}
 & P_{01}=1-s-t+P_{00}\\
 & P_{10}=P_{00}\\
 & P_{11}=-s+t+P_{00}\\
 & Q_0=2s-2P_{00}\\
 & Q_1=-2P_{00}
\end{split}
\end{align}

The last equation in Eq. (\ref{eq:sysfinal}) and the positivity of $P_{ij}$ and $Q_{\beta}$ together imply that $P_{00}=Q_{1}=0$. We are left with

\begin{align}\label{eq:constraints}
\begin{split}
& Q_1=P_{00}=P_{10}=0\\
& Q_0=2s\\
& P_{01}=1-s-t\\
& P_{11}=t-s
\end{split}
\end{align}

This solution explicitly defines several possible nonlocal ensembles which reduce to the two 1-extremal decompositions for mixed states belonging to the left triangle in Fig. \ref{fig1}, such as the point $S$. Indeed, Eq. (\ref{eq:constraints}) implies
\begin{align}\label{eq:constraints2}
\begin{split}
& q_{\alpha 1\delta}=0,\ \forall\alpha ,\delta\\
& p_{00kl}=p_{10kl}=0,\ \forall k,l
\end{split}
\end{align}
which implies that the local part of the nonlocal ensemble reduces to the boxes $S_{01}$ and $S_{11}$. The nonlocal part collapses to different boxes depending on Bob's input $y$. The first condition in Eq. (\ref{eq:constraints2}) implies only $\mathrm{PR}_{\alpha 0\delta}$ boxes are used, and these are defined through the algebraic relation
\begin{equation}
a=xy\oplus\alpha y\oplus\delta\oplus b
\end{equation}
reducing to $S_{00},S_{01}$ when $y=0$ and $S_{10},S_{11}$ when $y=1$. Therefore $y=0$ collapses the nonlocal ensemble to the upper triangle decomposition ($\epsilon_{10}=0$) in Fig. \ref{fig1}, while $y=1$ gives the lower triangle decomposition ($\eta_{00}=0$).

If Bob informs the Referee about his measurement choice and outcome in every round, the Referee knows which constituent state Alice holds. If the state in a given round is a product $S_{ij}\times S_{kl}$, then Bob's information serves no purpose, the Referee knows the constituent state Alice holds. If, on the other hand, the state is a particular $\mathrm{PR}_{\alpha 0\delta}$, then Alice's constituent state reduces to $S_{y,\alpha y+\delta+b}$ and the Referee only knows which one Alice holds if he is told $y$ and $b$.

\section{Conclusion}

Our investigation of the idea of steering in no-signalling theories has led us to discover fundamental things about the meaning of steering. A first result is that in quantum mechanics, ``steering'' involves a collection of different aspects; going to more general theories we learn that these aspects that were collated together in quantum steering, need to be separated. In particular, in quantum mechanics we can implement general POVMs by using an ancilla. This allows one to overcome the limitations deriving from the dimension of the Hilbert space of the system and allows strong steering.  However, the possibility of performing POVMs by using ancillas has basically nothing to do with the concept of steering; it is an independent property of quantum mechanics. The study of steering in generalised theories, clearly exposes this difference: In quantum mechanics performing POVMs with ancillas is possible due to the possibility of making entangling measurements between the system and ancilla, a possibility that doesn't exist in generalised theories. However steering is possible by preparing from the beginning a two partite state in which the steerer (Bob) has a larger ``dimensional'' space of states (i.e. measurements have a large enough number of independent possible results). 

At the same time, when the steerer's system has only limited ``dimension'', we see that the possibility of steering still persists. We introduced the concept of ``blind steering'' which, in some sense, is the most basic form of steering.  At its most basic level, all that we require from the steerer is to steer, not necessarily know in to what states he steers the system, as long as someone else, a Referee, can verify that the correct steering occurred. 

We conclude by posing two open problems. Firstly, it is known that some particular no-signalling theories beyond quantum mechanics allow for ``entangling'' measurements (so called ``couplers''), see eg \cite{Skrzypczyk2009}. It would be interesting to study steering in these particular theories. Secondly, and since blind steering is present in both no-signalling theories and quantum mechanics, an exciting topic for further research would be to study the phenomenon in scenarios which combine the two kinds of theories, such as the ones considered in \cite{Sainz2020}.

\acknowledgements{This work is supported by the Swiss National Science Foundation via the Mobility Fellowship P2GEP2 188276 and the NCCR-SwissMap.}

\bibliography{blind}

\end{document}